\documentstyle[12pt,epsf,axodraw]{article}

\newcommand{\beq}{\begin{equation}}
\newcommand{\eeq}{\end{equation}}
\def\bea{\begin{eqnarray}}
\def\eea{\end{eqnarray}}
\def\nn{\nonumber}

\begin{document}

\newcommand{\sheptitle}{Sufficiently Small 
                   $\bar{\theta}$ in $SU(3)^3 \times S_3$ Unification Model}
\newcommand{\shepauthor}{K. Chalut, ~H. Cheng, 
                        ~P.H. Frampton, ~K. Stowe ~and ~T. Yoshikawa}
\newcommand{\shepaddress}{Department of Physics and Astronomy,
                       University of North Carolina,\\
                       Chapel Hill, NC 27599-3255.}

\date{\today}

\begin{titlepage}
\begin{flushright}
hep-ph/0204074\\
IFP-807-UNC\\
\today
\end{flushright}
\vspace{.1in}
\begin{center}
{\large{\bf \sheptitle}}
\bigskip \medskip \\
\shepauthor \\
\mbox{} \\
{\it \shepaddress} \\
\vspace{.5in}

\bigskip \end{center} \setcounter{page}{0}
%\shepabstract
\begin{abstract}
Since CP violation in weak decays is successfully described by the
KM mechanism, the strong CP problem cannot easily be accommodated. 
This leads us to reconsider
the issue. If the axion and massless up quark are abandoned, we must
extend the standard model. Extension to $SU(3)^3 \times S_3$ unification
leads to the following situation: {\it if} CP is a high-energy
symmetry and the appropriate
symmetry-breaking hierarchy of scales is in place, 
then the $\bar{\theta}$ parameter
of the QCD sub-theory is guaranteed to be sufficiently
small. We find $\bar{\theta} < 10^{-11}$ while
the empirical limit from the neutron electric dipole
moment requires only that $\bar{\theta} < 1.3 \times 10^{-10}$.
\end{abstract}

\vspace*{6cm}

\begin{flushleft}
\hspace*{0.9cm} \begin{tabular}{l} \\ \hline {\small Emails:
hawcheng@hotmail.com, ~chalut@phy.duke.edu, ~frampton@physics.unc.edu,}\\
{\small stowek@physics.unc.edu, ~tadashi@physics.unc.edu} \\
\end{tabular}
\end{flushleft}

\end{titlepage}

\newpage

\bigskip
\bigskip
\bigskip

\section{Introduction}

\bigskip
\bigskip

\indent Within the last year, we have learned a great deal
about the nature of CP violation.
Results from the B Factories \cite{BF1,BF2} have shown unambiguously
that there is a large CP asymmetry in the decay mode
$( B^0, \bar{B}^0 ) \rightarrow \Psi K_S$. The resultant value for
the parameter $\sin 2 \beta$ is $\sin 2 \beta = 0.79 \pm 0.10$ which
is eight standard deviations from zero and fully consistent
with the prediction of the standard model \cite{SM}.

\bigskip

From the viewpoint of fundamental theory, the data suggest that
explicit CP violation is at work and disfavor models based on 
soft CP violation
\cite{SPCPV} which generically, although not universally, predicted
a value of $\sin 2\beta$ too small to be observable in the present 
B Factories.

\bigskip

One advantage of soft CP violation was that it offered a natural
solution of the strong CP problem: the value of $\bar{\theta}$
vanished at tree-level by virtue of the assumption that CP
was an exact symmetry of the lagrangian and radiative corrections
then contributed a sufficiently small value of $\bar{\theta}$ to be
acceptable.

\bigskip

It now appears that this approach is disfavored and therefore one 
must reconsider
the strong CP problem. Two other well-known approaches are, in our opinion,
equally as disfavored as soft CP violation:

\bigskip

\noindent (i) The invisible axion suffers from unacceptable fine-tuning
when gravitational effects are considered\cite{KKH++}, actually a greater
degree of fine-tuning than required to solve the original issue.

\bigskip

\noindent (ii) The massless up quark is strongly disfavored by the most
recent analysis of lattice gauge theories\cite{lattice}

\bigskip

Thus we are led to reconsider the strong CP problem. In particular, it
requires some extension of the standard model. The Left-Right\cite{LR} 
model has been much
studied from this point of view\cite{LRCP} and
the conclusion is that 
additional discrete symmetries must be added to do the job.

\bigskip

In the present work we present a unified extension of the standard
model in which, {\it provided} CP is an exact
high-energy symmetry of the lagrangian ${\cal L}$
and the hierarchy of symmetry-breaking
scales is taken care
of properly, the strong CP problem is resolved.

\newpage

\bigskip
\bigskip

\section{The Model}

\bigskip
\bigskip

We consider unification of the standard model in the
semi-simple gauge group $SU(3)^3 $ as suggested in \cite{RGG}.
In particular we follow the notation of \cite{CW} for
the fields and subscripts.

The
unification group is
$G = SU(3)_C \times SU(3)_L \times SU(3)_R \times S_3$
where $SU(3)_C$ is color, $SU(3)_L$ contains the 
$SU(2)_L$ of electroweak interactions as in \cite{331}, and the remaining $U(1)_Y$
is distributed between $SU(3)_L$ and $SU(3)_R$.  In addition, there is a
permutation $S_3$ symmetry relating the three $SU(3)$ gauge groups. This $S_3$ symmetry 
includes a cyclic $Z_3$ subgroup and three pair switching permutations. 
One of them has a role of parity symmetry. 

The scalar bosons and each family of fermions is assigned to a 27 dimensional 
representation of $G$.
They transform under $G$ as the representation
$\Psi = \psi_L(3 , \bar 3, 1) + \psi_R(\bar 3, 1, 3) 
+ \psi_\ell (1, 3, \bar 3)$, where 
\bea
\psi_L(3,\bar{3},1)&:& U_C \left( \begin{array}{ccc}
u_1 & ~d_1~ & B_1 \\ u_2 & d_2 & B_2 \\ 
u_3 & d_3 & B_3 
\end{array}
\right) \overline{V}_L, \\
\psi_R(\bar{3},1,3)&:& W_R \left( \begin{array}{ccc}
\bar{u}_1&\bar{u}_2&\bar{u}_3 \\ \bar{d}_1&\bar{d}_2&\bar{d}_3 \\ 
\bar{B}_1 & \bar{B}_2 & \bar{B}_3 
\end{array} 
\right) \overline{U}_C, \\ 
\psi_\ell(1,3,\bar{3})&:& V_L \left( \begin{array}{ccc}
E^0&E^-&e^- \\ E^+&\bar{E}^0&\nu \\ 
e^+ & N^0 & \bar{N}^0 
\end{array}
\right) \overline{W}_R. 
\eea
$U,W,$ and $V$ are group elements of the three $SU(3)$ groups. 
The left-handed quarks and anti-right-handed quarks will be found in $\psi_L$
and $\psi_R$ respectively, and the leptons are found inside $\psi_\ell$.  
There are a heavy weak singlet quark B with charge $-1/3$ and a heavy lepton
doublet $E^0$ and $E^+$ with charge $0$ and $1$. In addition, there is 
a neutral chiral state $N^0$ which may be called neutretto.  

The scalars will acquire vacuum expectation values (VEV's) which are arranged as
\begin{equation}
\left< \phi_{\ell}(1,3,\bar{3}) \right> = \left( \begin{array}{ccc}
u&0&0 \\ 0&u&u \\ 
0 & w & v
\end{array}
\right)
\end{equation}
The `constants' $u$, $v$, and $w$ here merely
represent orders of magnitude, rather than specific values. 
For the scales we assume, as in \cite{CW}, that the hierarchy $v \gg w \gg u$
is incorporated in the $SU(3)^3 \times S_3$ model with
whatever fine-tuning is necessary to accomplish it. Our
aim here is to show only that the strong
CP parameter $\bar{\theta}$ is under control
without {\it further} fine tuning.

\bigskip 
 
The unification scale $v$ breaks $SU(3)^3 \times S_3$ down to 
$SU(3)_C \times SU(2)_L \times SU(2)_R \times U(1)_{B-L} \times P$.  
At the scale $w$, 
the symmetry break down to the standard model 
$SU(3)_C \times SU(2)_L \times SU(1)_Y$.
Finally, the scale $u$ accomplishes the electroweak breaking.  
If there is only one scalar field, the
VEV can always be diagonalized, so that it is impossible to have the standard
model at an intermediate scale.  Hence it is necessary to assume at least two
{\bf 27}'s, and, in the interest of economy, we will assume exactly two {\bf 27}'s.  
Hence, we will assume two {\bf 27}'s of scalars for Higgs fields, 
three {\bf 27}'s of fermions for the generations and one {\bf 24} of gauge boson 
in this model. 

\bigskip
\bigskip

We must treat one of the pair switching permutations in the $S_3$ symmetry as parity
symmetry\cite{CW}.  Let us write down the action of ${\bf P}$
on all the fields:
\begin{eqnarray}
\psi^A_\ell(\vec x, t)  \rightarrow  \psi^{A\dagger}_\ell(- \vec x, t), 
& \phi^i_\ell(\vec x, t)  \rightarrow  \phi^{i\dagger}_\ell(- \vec x, t), &
C^\mu_a(\vec x, t)  \rightarrow  C_{a\mu} (-\vec x, t) \nonumber \\
\psi^A_L(\vec x, t)  \rightarrow  \psi^{A\dagger}_R(- \vec x, t), &
\phi^i_L(\vec x, t)  \rightarrow \phi^{i\dagger}_R(- \vec x, t), &
L^\mu_a(\vec x, t)  \rightarrow  R_{a\mu} (-\vec x, t) \nonumber \\ 
\psi^A_R(\vec x, t)  \rightarrow  \psi^{A\dagger}_L(- \vec x, t), &
\phi^i_R(\vec x, t)  \rightarrow \phi^{i\dagger}_L(- \vec x, t), &
R^{\mu}_{a}(\vec x, t)  \rightarrow  L_{a\mu} (-\vec x, t)
\end{eqnarray} 
where $C$, $R$, and $L$ are the gauge fields, the lowering of the index $^\mu$
to $_\mu$ indicates reversal of the spatial components, $A=1,2,3$ is a family
index and $i=1,2$ is a gauge boson index.  The daggers ($^\dagger$) represent
the
fact that not only are each component of these matrices complex conjugated,
but the $SU(3)_L$ and $SU(3)_R$ indices are exchanged as well.

The Yukawa couplings are given by
\begin{equation}
{\cal L}_{\rm Y} = 
- Z_3\{ f_{iAB} {\rm Tr}(\phi_\ell^i \psi_R^A \psi_L^B) + h.c.\}
\end{equation}
where $Z_3$ simply implies that we must include cyclic permutations to assure
the $Z_3$ symmetry is respected.  Under the symmetry ${\bf P}$, we can
relate the terms to their hermitian conjugates, so that $f_{iAB}^* = f_{iBA}$,
or, thinking of these as matrices, $f_i^\dagger = f_i$.  If the scalar
VEV's are real (proved below), this will result in Hermitian quark mass matrices, or
$M_q^\dagger=M_q$.  Since the determinant of a Hermitian matrix is real, this in turn 
would result in vanishing $\bar\theta$ at tree level. It remains only to prove the
$<\phi_{\ell}^i>$ are real.

\bigskip

\section{Reality of $\langle \phi_\ell^i \rangle$ at tree level} 
The scalar potential responsible for the symmetry
breaking involves only the $\phi_\ell$ portions which acquire  VEV's.
The portion of the scalar potential that is relevant is given by
\def\pl{\phi_\ell}
\begin{eqnarray}
-{\cal L}_{\ell} = & m^2_{ij} {\rm Tr}(\pl^{i\dagger} \pl^j)
 + \{\gamma_{ijk} \epsilon_{\alpha\beta\gamma} \epsilon^{\delta\sigma\rho} 
{\pl^i}_\delta^\alpha {\pl^j}_\rho^\beta {\pl^k}_\sigma^\gamma \; + \; h.c. \} \nonumber \\ 
& + \lambda_{ijkm} {\rm Tr}(\pl^{i\dagger} \pl^j) {\rm Tr}(\pl^{k\dagger}
\pl^m)
+ \eta_{ijkm} {\rm Tr}(\pl^{i\dagger} \pl^j \pl^{k\dagger} \pl^m). 
\end{eqnarray}
All the coefficients in this potential are real because of  
parity symmetry and the hermiticity.  
Hermiticity and parity imply
\bea
m_{ij}^2 = m_{ji}^{2*} = m_{ij}^{2*}, ~~~ \gamma_{ijk} = \gamma_{ijk}^*,~~~
\lambda_{ijkm} = \lambda_{jimk} = \lambda_{jimk}^*,
\eea
\bea
\eta_{ijkm} = \eta_{mijk} = \eta_{kmij} =\eta_{jkmi} = 
\eta_{mkji}^* ,
\label{HPdemand}
\eea 
{}From these conditions we can find $m_{ij}, \gamma_{ijk}$ and 
$\lambda_{ijkm}$ must be real values.  $\eta_{ijkm}$ is also real 
because the indices take on only the values 1 or 2.  
Thus the whole potential ${\cal L}_{\ell}$ is real.  

%In the next section, we will show that, in fact, the VEVs
%of $\phi_{\ell}$ are real, as already hinted by the reality
%of ${\cal L}_{\ell}$.

Under the condition that all the constants in the potential are real, 
using the degree of freedom of gauge symmetry (rotation), 
we can take the VEV as follows;
\begin{equation}
\left\langle \phi_{\ell}^1 \right\rangle = \left( \begin{array}{ccc}
u_1 e^{i\alpha_1}&0&0 \\ 0&u_2 &  0 \\ 
0 & 0 & v
\end{array}
\right)  ~~ \hbox{and} ~~  
\left\langle \phi_{\ell}^2 \right\rangle = \left( \begin{array}{ccc}
u_3 e^{i\alpha_3} & 0 &0 \\ 0& u_4 ~e^{i\alpha_4 } &  u_5 ~e^{i\alpha_5 } \\ 
0 & w & x ~e^{i \alpha_x }  
\end{array}
\right) 
\end{equation}
where $u_i (i=1,2,3,4,5), v, w$ and $ x $ are real values.   

As a minimum case to realize the SM, we can assume 
that $u_1,u_4,u_5,x =0$. Then,
\begin{equation}
\left\langle \phi_{\ell}^1 \right\rangle = \left( \begin{array}{ccc}
0 & 0 & 0 \\ 0&u_2 &  0 \\ 
0 & 0 & v
\end{array}
\right)  ~~ \hbox{and} ~~ 
\left\langle \phi_{\ell}^2 \right\rangle = \left( \begin{array}{ccc}
u_3 e^{i\alpha_3}  & 0 & 0 \\ 0& 0 & 0 \\ 
0 & w & 0  
\end{array}
\right) 
\label{VEV}
\end{equation}  
The remaining phase of the VEV is only $\alpha_3$
The stationary condition for the remaining phase $\alpha_3$ appear 
in only the cubic interactions because
the other terms are completely real.   
\bea
\frac{d V}{d \alpha_3 } = -12 \gamma_{112} u_2 u_3 v \sin\alpha_3 =0. 
\eea  
So this condition shows $\alpha_3 = 0 $ (or $\pi $ ) 
and all VEVs are real 
in this minimum case. 
This confirms the earlier assertion
that $\bar{\theta} = 0$ at tree level,
and it remains to confirm whether the loop
corrections to $\bar{\theta}$
are sufficiently small to satisfy
the empirical constraint $\bar{\theta} < 1.3 \times 10^{-10}$\cite{EDM}.

\bigskip
\bigskip
\section{The contribution from loop diagrams}
Some possible CP violating effects remain  
in interactions among $\phi_L^i(3,\bar{3},1), \phi_R^i(\bar{3},1,3)$ 
and $\phi_\ell^i(1,3,\bar{3})$. 
Here we consider only such terms which can contribute 
to the imaginary part of quark mass terms at either one or two loop level :
\bea
V_{CP violating} & = & A^L_{ijkm} Tr(\phi_\ell^{i\dagger} \phi_\ell^j) Tr(\phi_L^{k \dagger} \phi_L^m) 
+A^R_{ijkm} Tr(\phi_\ell^{i\dagger} \phi_\ell^j) Tr(\phi_R^{k \dagger} \phi_R^m)
  \nonumber \\
&+&B^L_{ijkm} Tr(\phi_\ell^i \phi_\ell^{j\dagger} \phi_L^{k\dagger} \phi_L^{m}) 
+B^R_{ijkm} Tr(\phi_\ell^{i \dagger} \phi_\ell^{j} \phi_R^k \phi_R^{m \dagger})
  \nonumber \\
&+&C_{ijkm} (\phi_\ell^i)^\alpha_\rho (\phi_\ell^j)^\beta_\sigma 
         (\phi_L^{k\dagger})^x_\delta (\phi_R^{m\dagger})^\gamma_x 
          \varepsilon_{\alpha\beta\gamma} 
          \varepsilon^{\rho\sigma\delta} \nn \\
&+&D_{ikm} Tr(\phi_\ell^i \phi_R^k \phi_L^m )   +h.c. 
\label{CPV}
\eea  
where $i,j,k,m$ are Higgs scalar indices. 
Under {\bf P} invariance and hermiticity, 
we get the following conditions for each 
coupling constant,  
\bea
A^L_{ijkm} = A^{L~*}_{jimk} = A^{R~*}_{ijkm} = A^{R}_{jimk},
\label{ALR}\\ 
B^L_{ijkm} = B^{L~*}_{jimk} = B^{R~*}_{jimk} = B^R_{ijkm},
\label{BLR}\\ 
C_{ijkm} = C_{ijmk}^*, ~~ D_{ikm} = D_{imk}^*. 
\label{CDLR} 
\eea
$A^{L,R}$ and $B^{L,R}$ are real if $i=j$ and $k=m$. $C$ and $D$ are also real if 
$k=m$. However there is no constraint for the other cases 
$i \ne j$ or  $k \ne m$. 
So these constants are in general complex. This model has ${\bf P}$ symmetry between 
the scales $v$ and $w$ so that $\phi_L$ and $\phi_R$ are degenerate. 
Above the scale $w$ any imaginary contributions are cancelled and $\bar{\theta}$ vanishes
in all orders.  Below the scale $w$, $\bar{\theta}$ may have non zero value 
through loop diagrams.  The imaginary parts of these interaction 
contribute to the quark mass through Yukawa couplings
\bea
- {\cal L}_{\rm Y} = Z_3\left[ f_{iAB} 
      Tr\left(\phi_\ell^i\psi_R^A \psi_L^B \right) 
     + g_{iAB} Tr\left\{(\phi_\ell^i)^\alpha_\rho 
         (\psi_\ell^A)^\beta_\sigma (\psi_\ell^B)^\gamma_\delta 
          \varepsilon_{\alpha \beta \gamma }
          \varepsilon^{\rho \sigma \delta } \right\}  +h.c. ~\right].
\eea
The up-type quark masses arise from the VEV of $\phi_\ell^2$ and down-type 
quark masses come from $\phi_\ell^1$ if we choose the sets of 
VEV as Eq.(\ref{VEV}). 
These Yukawa couplings are proportional to the mass of quarks and leptons 
for each family. 
\bea
 f_{2AB} &=& \frac{m_{u^A}}{u} \hat{f}_{2AB} \\
 f_{1AB} &=& \frac{m_{d^A}}{u} \hat{f}_{1AB} \\
 g_{iAB} &=& \frac{m_{l}}{u} \hat{g}_{iAB}.  
\eea  
In the case of the coupling to charged Higgs, $\hat{f}_{iAB}$ includes 
the Kobayashi-Maskawa matrix. 
We investigate the loop effects to $\bar{\theta}$ 
through these interactions 
in the follow  subsections. 
 
\bigskip

\subsection{One loop}

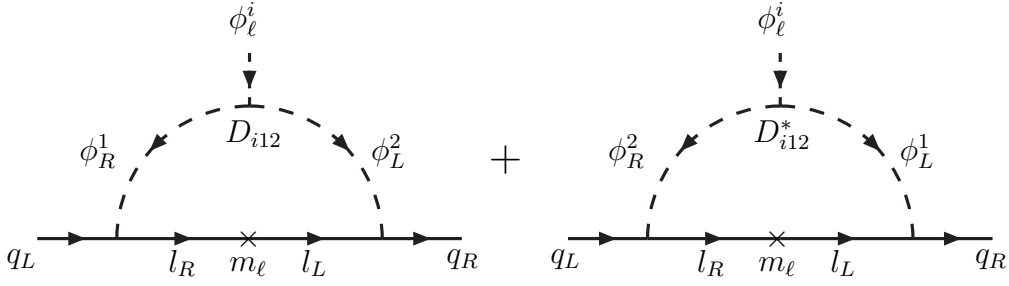
\begin{figure}[ht]
\begin{picture}(410,120)(0,0)
\SetWidth{1.2}
{
\ArrowLine(25,20)(55,20)
\ArrowLine(55,20)(105,20)
\ArrowLine(105,20)(155,20)
\ArrowLine(155,20)(185,20)
\DashArrowArc(105,20)(50,90,180){5}
\DashArrowArcn(105,20)(50,90,0){5}
\DashArrowLine(105,90)(105,70){5}
\ArrowLine(225,20)(255,20)
\ArrowLine(255,20)(305,20)
\ArrowLine(305,20)(355,20)
\ArrowLine(355,20)(385,20)
\DashArrowArc(305,20)(50,90,180){5}
\DashArrowArcn(305,20)(50,90,0){5}
\DashArrowLine(305,90)(305,70){5}
\Text(105,65)[t]{ $D_{i12}$}
\Text(305,65)[t]{ $D_{i12}^*$}
\Text(25,16)[rt]{$q_L$}
\Text(180,16)[lt]{$q_R$}
\Text(225,16)[t]{$q_L$}
\Text(380,16)[lt]{$q_R$}
\Text(80,16)[t]{$l_R$}
\Text(130,16)[t]{$l_L$}
\Text(280,16)[t]{$l_R$}
\Text(330,16)[t]{$l_L$}
\Text(105,20)[]{$\times $}
\Text(105,13)[t]{$m_\ell $}
\Text(305,20)[]{$\times $}
\Text(305,13)[t]{$m_\ell $}
\Text(50,50)[b]{$\phi_R^1$ }
\Text(160,50)[b]{$\phi_L^2$ }
\Text(105,100)[b]{$\phi_\ell^i$ }
\Text(250,50)[b]{$\phi_R^2$ }
\Text(360,50)[b]{$\phi_L^1$ }
\Text(305,100)[b]{$\phi_\ell^i$ }
\Text(205,50)[]{\bf \Large + }}
\end{picture}\\
\caption{ One-loop contribution to
$\bar{\theta}$ involving
$D_{ijk}$
and internal lepton.}
\label{one-loop1}
\end{figure}

There are contributions from one loop diagrams to the imaginary part of the quark mass 
from the Feynman graphs of Figs. \ref{one-loop1}, 
\ref{one-loop2},  
\ref{one-loop3},  
and \ref{one-loop4}.  
These one loop diagrams are proportional to 
\bea
\frac{1}{(4 \pi)^2 } \frac{1}{ M_{\phi_L^i}^2 - M_{\phi_R^j}^2}
           log\frac{M_{\phi_L^i}^2}{M_{\phi_R^j}^2}, 
\eea
where $M_\phi $ is the mass of the colored heavy higgs mass. 
Above the scale $w$ 
these masses became degenerate and the difference between $M_{\phi_L^i}$ 
and $M_{\phi_R^i}$ therefore arises 
from the difference of contribution at scale $w$ after 
breaking the ${\bf P }$ 
symmetry. In general, we can write the masses as follows:
\bea
M_{\phi_L^i}^2 &=& \alpha_L^i v^2 + \beta_L^i w^2 \\
M_{\phi_R^i}^2 &=& \alpha_L^i v^2 + \beta_R^i w^2 
\eea 
By the relations of Eq.(\ref{CDLR}) and the hermiticity of 
the Yukawa couplings, 
the imaginary part of the sum of the two diagrams in each of Figs. 
\ref{one-loop1}-\ref{one-loop4} is suppressed 
by a factor

\begin{figure}[ht]
\begin{picture}(410,120)(0,0)
\SetWidth{1.2}
{
\ArrowLine(25,20)(55,20)
\ArrowLine(55,20)(105,20)
\ArrowLine(105,20)(155,20)
\ArrowLine(155,20)(185,20)
\DashArrowArc(105,20)(50,90,180){5}
\DashArrowArcn(105,20)(50,90,0){5}
\DashArrowLine(80,100)(105,70){5}
\DashArrowLine(130,100)(105,70){5}
\ArrowLine(225,20)(255,20)
\ArrowLine(255,20)(305,20)
\ArrowLine(305,20)(355,20)
\ArrowLine(355,20)(385,20)
\DashArrowArc(305,20)(50,90,180){5}
\DashArrowArcn(305,20)(50,90,0){5}
\DashArrowLine(280,100)(305,70){5}
\DashArrowLine(330,100)(305,70){5}
\Text(105,65)[t]{ $C_{ij12}$}
\Text(305,65)[t]{ $C_{ij12}^*$}
\Text(25,16)[rt]{$q_L$}
\Text(180,16)[lt]{$q_R$}
\Text(225,16)[t]{$q_L$}
\Text(380,16)[lt]{$q_R$}
\Text(80,16)[t]{$l_R$}
\Text(130,16)[t]{$l_L$}
\Text(280,16)[t]{$l_R$}
\Text(330,16)[t]{$l_L$}
\Text(105,20)[]{$\times $}
\Text(105,13)[t]{$m_\ell $}
\Text(305,20)[]{$\times $}
\Text(305,13)[t]{$m_\ell $}
\Text(50,50)[b]{$\phi_R^1$ }
\Text(160,50)[b]{$\phi_L^2$ }
\Text(75,110)[b]{$\phi_\ell^i$ }
\Text(140,110)[b]{$\phi_\ell^j$ }
\Text(250,50)[b]{$\phi_R^2$ }
\Text(360,50)[b]{$\phi_L^1$ }
\Text(275,110)[b]{$\phi_\ell^i$ }
\Text(340,110)[b]{$\phi_\ell^j$ }
\Text(205,50)[]{\bf \Large + }}
\end{picture}\\
\caption{ One-loop contribution to
$\bar{\theta}$ involving
$C_{ijk\ell}$
and internal lepton.}
\label{one-loop2}
\end{figure}

\begin{center}
\begin{figure}[ht]
\begin{picture}(410,120)(0,0)
\SetWidth{1.2}
{
\ArrowLine(25,20)(55,20)
\ArrowLine(105,20)(55,20)
\ArrowLine(155,20)(105,20)
\ArrowLine(155,20)(185,20)
\DashArrowArc(105,20)(50,90,180){5}
\DashArrowArcn(105,20)(50,90,0){5}
\DashArrowLine(105,90)(105,70){5}
\ArrowLine(225,20)(255,20)
\ArrowLine(305,20)(255,20)
\ArrowLine(355,20)(305,20)
\ArrowLine(355,20)(385,20)
\DashArrowArc(305,20)(50,90,180){5}
\DashArrowArcn(305,20)(50,90,0){5}
\DashArrowLine(305,90)(305,70){5}
\Text(105,65)[t]{ $D_{i12}$}
\Text(305,65)[t]{ $D_{i12}^*$}
\Text(25,16)[rt]{$q_L$}
\Text(180,16)[lt]{$q_R$}
\Text(225,16)[t]{$q_L$}
\Text(380,16)[lt]{$q_R$}
\Text(80,16)[t]{$q_L$}
\Text(130,16)[t]{$q_R$}
\Text(280,16)[t]{$q_L$}
\Text(330,16)[t]{$q_R$}
\Text(105,20)[]{$\times $}
\Text(105,13)[t]{$m_q $}
\Text(305,20)[]{$\times $}
\Text(305,13)[t]{$m_q $}
\Text(50,50)[b]{$\phi_L^1$ }
\Text(160,50)[b]{$\phi_R^2$ }
\Text(105,100)[b]{$\phi_\ell^i$ }
\Text(250,50)[b]{$\phi_L^2$ }
\Text(360,50)[b]{$\phi_R^1$ }
\Text(305,100)[b]{$\phi_\ell^i$ }
\Text(205,50)[]{\bf \Large + }}
\end{picture}\\
\caption{ One-loop contribution to
$\bar{\theta}$ involving
$D_{ijk}$
and internal quark. }
\label{one-loop3}
\end{figure}
\end{center}

\begin{center}
\begin{figure}[ht]
  \begin{picture}(410,120)(0,0)
\SetWidth{1.2}
{
\ArrowLine(25,20)(55,20)
\ArrowLine(105,20)(55,20)
\ArrowLine(155,20)(105,20)
\ArrowLine(155,20)(185,20)
\DashArrowArc(105,20)(50,90,180){5}
\DashArrowArcn(105,20)(50,90,0){5}
\DashArrowLine(80,100)(105,70){5}
\DashArrowLine(130,100)(105,70){5}
\ArrowLine(225,20)(255,20)
\ArrowLine(305,20)(255,20)
\ArrowLine(355,20)(305,20)
\ArrowLine(355,20)(385,20)
\DashArrowArc(305,20)(50,90,180){5}
\DashArrowArcn(305,20)(50,90,0){5}
\DashArrowLine(280,100)(305,70){5}
\DashArrowLine(330,100)(305,70){5}
\Text(105,65)[t]{ $C_{ij12}$}
\Text(305,65)[t]{ $C_{ij12}^*$}
\Text(25,16)[rt]{$q_L$}
\Text(180,16)[lt]{$q_R$}
\Text(225,16)[t]{$q_L$}
\Text(380,16)[lt]{$q_R$}
\Text(80,16)[t]{$q_L$}
\Text(130,16)[t]{$q_R$}
\Text(280,16)[t]{$q_L$}
\Text(330,16)[t]{$q_R$}
\Text(105,20)[]{$\times $}
\Text(105,13)[t]{$m_q $}
\Text(305,20)[]{$\times $}
\Text(305,13)[t]{$m_q $}
\Text(50,50)[b]{$\phi_L^1$ }
\Text(160,50)[b]{$\phi_R^2$ }
\Text(75,110)[b]{$\phi_\ell^i$ }
\Text(140,110)[b]{$\phi_\ell^j$ }
\Text(250,50)[b]{$\phi_L^2$ }
\Text(360,50)[b]{$\phi_R^1$ }
\Text(275,110)[b]{$\phi_\ell^i$ }
\Text(340,110)[b]{$\phi_\ell^j$ }
\Text(205,50)[]{\bf \Large + }}
\end{picture}\\
\caption{ One-loop contribution to
$\bar{\theta}$ involving
$C_{ijk\ell}$
and internal quark.}
\label{one-loop4}
\end{figure}
\end{center}

\newpage

\bea
\frac{1}{M_{\phi_R^1}^2 - M_{\phi_L^2}^2 } 
- \frac{1}{M_{\phi_L^1}^2 - M_{\phi_R^2}^2 } = \left\{ 
\frac{\beta_L^1-\beta_L^2 -\beta_R^1+\beta_R^2}{(\alpha_L^1 - \alpha_L^2)^2 }
 \right\} \frac{w^2}{v^4}   \propto  \frac{w^2}{v^4}.
\eea 
Thus, all one-loop diagrams have a suppression factor $\frac{w^2}{v^2}$ in
$\bar{\theta}$. 

The largest contribution to $\bar{\theta }$ comes from the loop 
correction to the third family quark mass
because the Yukawa coupling and the mass are much larger 
than the other family.
When the scalar has the largest VEV $v$, the    
estimates of these one loop diagrams are  
\bea
\bar{\theta} = \frac{1}{m}{\rm Im}[Fig.1] 
 &\sim & v ~{\rm Im}[D_{112}] 
    \hat{f}_{133}\hat{f}_{233} \frac{1}{(4 \pi)^2} \left(\frac{w}{v}\right)^2 
              \frac{m_\tau}{v^2} (\frac{m_t m_b}{u^2}) \frac{1}{m_t} \\
\bar{\theta} = 
      \frac{1}{m}{\rm Im}[Fig.2] &\sim & u v \hat{f}_{133}\hat{f}_{233} 
        \frac{1}{(4 \pi)^2} \left(\frac{w}{v}\right)^2 
              \frac{m_E}{v^2} (\frac{m_t m_b}{u^2})
                   (\frac{{\rm Im}[C_{1212}]}{m_b}
                   +\frac{{\rm Im}[C_{1112}]}{m_t} ) 
\eea
where $D_{112}$ has a mass dimension and 
it will be at largest order of scale $v$. $m_E$ is the mass of heavy lepton 
which is proportional to the lepton Yukawa coupling. In the third family
case, 
\bea
 m_E = \hat{g}_{133} \frac{m_\tau}{u} v . 
\eea
In Fig. \ref{one-loop1}, the contribution to bottom quark mass is negligible 
because it proportional to the neutrino mass. So largest contribution 
comes from the loop correction to bottom mass of Fig. \ref{one-loop2}.     
If the size of the suppression factor $\left(\frac{w}{v}\right)^2$ is 
$O(10^{-7})$ as estimated in ref.\cite{CW}, the largest contribution will be 
$10^{-11}$ except for further unknown parameters expected to be less than one. 

Figs.\ref{one-loop3} and \ref{one-loop4} give contributions to $\bar{\theta}$ 
smaller than the previous two because these diagrams contain leptonic 
Yukawa coupling $g_{133}$, necessarily smaller than its quark counterpart. 

\bigskip
\bigskip

Hence, in summary, the contributions from all one-loop diagrams to $\bar{\theta}$ 
add to a value smaller than $10^{-11}$.  

\newpage

\bigskip
\bigskip

\subsection{Two loops}
The effective complex quartic interactions of $\phi_\ell^i$ will appear 
through loops of colored scalars  as illustrated by the
four Feynman graphs of Fig. \ref{mixing} which involve
the $A^{L,R}_{ijkl}$ (recall that $A^R$ is related to $A^{L*}$)
quartic
couplings of Eq.(\ref{CPV}). There are similar
graphs for the $B^{L,R}_{ijkl}$ and $C_{ijkl}$ quartic
couplings, as well as a similar box diagram
involving the cubic term $D_{ijk}$ in Eq.(\ref{CPV}).  
The loop diagram in Fig. \ref{mixing} is proportional to $A^L_{ijxy} A^L_{klyx}$ (or 
$A^R_{ijxy} A^R_{klyx}$), where
$x,y$ show the indices of the colored scalars. If $i=l$ and $j=k$, 
this combination will be real because of the relations of the coefficients,
$A^{L,R}_{ijxy} = A^{L,R~*}_{ijyx}$ in 
Eq. (\ref{ALR}).  The case of $i=j, k=l$ is also real by adding 
the exchanged diagrams between the indices of the colored scalars 
$x$ and $y$. 
\bea
{\rm Im}[ A^L_{iixy} A^L_{kkyx} +  A^L_{iiyx} A^L_{kkxy} ] 
= {\rm Im}[ A^L_{iixy} A^L_{kkyx} +  A^{L*}_{iixy} A^{L*}_{kkyx} ] 
= 0  
\eea
Other combinations with at least one index different 
are also cancelled by adding the diagrams of $\phi_R$ as in 
Fig. \ref{mixing}. The difference between the diagrams of $\phi_L$ and 
$\phi_R$ comes only from mass splittings. Hence the remaining
imaginary part of the effective quartic interactions is suppressed 
by a factor $\left(\frac{w}{v}\right)^2 $. The effective couplings 
arising from $B^{L,R}_{ijkm}, C_{ijkm} $ and $D_{ikm}$ interactions 
are also suppressed by the same factor.

Two-loop diagrams which illustrate such contributions 
to $\bar{\theta} $ from effective quartic interactions are shown in 
Figs. \ref{two-loop1}-\ref{two-loop2}.  In Fig.\ref{two-loop1}, 
there are possible $A^{L,R}_{ijkl}$ and, separately,
$B^{L,R}_{ijkl}$ contributions
but the largest contribution comes from $B^{L,R}_{ijkm} $ interactions 
with one large VEV $v$. We estimate: 
\bea
\bar{\theta}_{Fig.\ref{two-loop1}} \sim
   u v ~{\rm Im}[B_{22xy}^L B_{12yx}^L]
   \frac{1}{(4 \pi)^4} \left( \frac{w}{v} \right)^2 \frac{m_B}{v^2} 
    \left(\hat{f}_{233}^2\frac{m_t^2}{u^2} \frac{1}{m_t} 
         +\hat{f}_{133}\hat{f}_{233}\frac{m_t m_b}{u^2} \frac{1}{m_b}
     \right),  
\eea
where $m_B$ is the heavy quark mass  
$m_B =  \frac{m_b}{u} v $ coming from the Yukawa coupling
to $\phi^1_\ell$.  So if the factor $\left( \frac{w}{v} \right)^2$ 
is $O(10^{-7})$, the size will become $O(10^{-13})$ except for unknown 
parameters expected to be smaller than one. 

The contribution from Fig.\ref{two-loop2} is estimated as: 
\bea
\bar{\theta}_{Fig.\ref{two-loop2}} \sim
   u v 
   \frac{1}{(4 \pi)^4} \left( \frac{w}{v} \right)^2 \frac{m_B}{v^2} 
    \frac{m_t^2}{u^2} \hat{f}_{233}^2
    \left({\rm Im}[C_{12xy} C_{22yx}] \frac{1}{m_t} + 
          {\rm Im}[C_{11xy} C_{22yx}] \frac{1}{m_b}
     \right).  
\eea 
The contribution is largest for the bottom quark mass but even for
it the contribution to $\bar{\theta}$ will be 
smaller than $10^{-11}$. 

Further examples of two-loop contributions are depicted
in Figs. \ref{two-loop3}, \ref{two-loop4}, \ref{two-loop5}.  
All such other two loop contributions to $\bar{\theta}$ are smaller than 
$10^{-13}$.

\begin{center}
\begin{figure}[htp]
\begin{picture}(410,140)(0,0)
\SetWidth{1.2}
{
\DashArrowArc(55,70)(30,-90,90){5}
\DashArrowArc(55,70)(30,90,270){5}
\DashArrowLine(30,20)(55,40){5}
\DashArrowLine(55,40)(80,20){5}
\DashArrowLine(30,120)(55,100){5}
\DashArrowLine(55,100)(80,120){5}
\DashArrowArc(255,70)(30,-90,90){5}
\DashArrowArc(255,70)(30,90,270){5}
\DashArrowLine(230,20)(255,40){5}
\DashArrowLine(255,40)(280,20){5}
\DashArrowLine(230,120)(255,100){5}
\DashArrowLine(255,100)(280,120){5}
\Text(55,95)[t]{ $A_{ij12}^{L}$}
\Text(255,95)[t]{ $A_{ij12}^{L*}$}
\Text(37,70)[b]{$\phi_L^1$ }
\Text(75,70)[b]{$\phi_L^2$ }
\Text(25,10)[b]{$\phi_\ell^l$ }
\Text(90,10)[b]{$\phi_\ell^k$ }
\Text(25,120)[]{$\phi_\ell^j$ }
\Text(90,120)[]{$\phi_\ell^i$ }
\Text(237,70)[b]{$\phi_R^1$ }
\Text(275,70)[b]{$\phi_R^2$ }
\Text(225,10)[b]{$\phi_\ell^l$ }
\Text(290,10)[b]{$\phi_\ell^k$ }
\Text(225,120)[]{$\phi_\ell^j$ }
\Text(290,120)[]{$\phi_\ell^i$ }
\Text(105,70)[]{\bf \Large + }
\Text(205,70)[]{\bf \Large + }
\DashArrowArc(155,70)(30,-90,90){5}
\DashArrowArc(155,70)(30,90,270){5}
\DashArrowLine(130,20)(155,40){5}
\DashArrowLine(155,40)(180,20){5}
\DashArrowLine(130,120)(155,100){5}
\DashArrowLine(155,100)(180,120){5}
\DashArrowArc(355,70)(30,-90,90){5}
\DashArrowArc(355,70)(30,90,270){5}
\DashArrowLine(330,20)(355,40){5}
\DashArrowLine(355,40)(380,20){5}
\DashArrowLine(330,120)(355,100){5}
\DashArrowLine(355,100)(380,120){5}
\Text(155,95)[t]{ $A_{ji12}^{L*}$}
\Text(355,95)[t]{ $A_{ji12}^{L}$}
\Text(137,70)[b]{$\phi_L^2$ }
\Text(175,70)[b]{$\phi_L^1$ }
\Text(125,10)[b]{$\phi_\ell^l$ }
\Text(190,10)[b]{$\phi_\ell^k$ }
\Text(125,120)[]{$\phi_\ell^j$ }
\Text(190,120)[]{$\phi_\ell^i$ }
\Text(337,70)[b]{$\phi_R^2$ }
\Text(375,70)[b]{$\phi_R^1$ }
\Text(325,10)[b]{$\phi_\ell^l$ }
\Text(390,10)[b]{$\phi_\ell^k$ }
\Text(325,120)[]{$\phi_\ell^j$ }
\Text(390,120)[]{$\phi_\ell^i$ }
\Text(305,70)[]{\bf \Large + }}
\end{picture}\\
\caption{CP-violating quartic interaction induced by
$A^{L,R}_{ijk\ell}$ couplings. }
\label{mixing}
%\end{figure}
%\end{center}

%\begin{center}
%\begin{figure}[h]
\begin{picture}(410,140)(0,0)
\SetWidth{1.2}
{
\DashArrowArc(205,70)(30,0,180){5}
\DashArrowArc(205,70)(30,180,360){5}
\DashArrowLine(155,120)(175,70){5}
\DashArrowLine(175,70)(155,20){5}
\DashArrowLine(255,120)(235,70){5}
\DashArrowLine(235,70)(255,20){5}
\ArrowLine(125,20)(155,20)
\ArrowLine(155,20)(205,20)
\ArrowLine(205,20)(255,20)
\ArrowLine(255,20)(285,20)
\Text(125,16)[rt]{$q_L$}
\Text(280,16)[lt]{$q_R$}
\Text(180,16)[t]{$q_R$}
\Text(230,16)[t]{$q_L$}
\Text(155,40)[b]{$\phi_\ell^i$ }
\Text(265,40)[b]{$\phi_\ell^k$ }
\Text(155,100)[b]{$\phi_\ell^j$ }
\Text(265,100)[b]{$\phi_\ell^l$ }
\Text(205,50)[b]{$\phi_{L,R}^x$ }
\Text(205,108)[b]{$\phi_{L,R}^y$ }
\Text(205,20)[]{$\times $ }}
\end{picture}\\
\caption{ Two-loop contribution to $\bar{\theta}$
involving $A^{L,R}_{ijkl}$ 
or $B^{L,R}_{ijk\ell}$ couplings.}
\label{two-loop1}
%\end{figure}
%\end{center}

%\begin{center}
%\begin{figure}[ht]
\begin{picture}(410,140)(0,0)
\SetWidth{1.2}
{
\DashArrowArc(205,70)(30,90,270){5}
\DashArrowArcn(205,70)(30,90,-90){5}
\DashArrowLine(175,120)(205,100){5}
\DashArrowLine(205,40)(155,20){5}
\DashArrowLine(235,120)(205,100){5}
\DashArrowLine(205,40)(255,20){5}
\ArrowLine(125,20)(155,20)
\ArrowLine(155,20)(205,20)
\ArrowLine(205,20)(255,20)
\ArrowLine(255,20)(285,20)
\Text(125,16)[rt]{$q_L$}
\Text(280,16)[lt]{$q_R$}
\Text(180,16)[t]{$q_R$}
\Text(230,16)[t]{$q_L$}
\Text(165,30)[b]{$\phi_\ell^i$ }
\Text(255,30)[b]{$\phi_\ell^j$ }
\Text(170,120)[b]{$\phi_\ell^k$ }
\Text(250,120)[b]{$\phi_\ell^m$ }
\Text(250,70)[b]{$\phi_{L}^x$ }
\Text(165,70)[b]{$\phi_{R}^y$ }
\Text(205,20)[]{$\times $ }
\Text(209,90)[]{$C_{kmxy}$ }
}
\end{picture}\\
\caption{ Two-loop contribution to $\bar{\theta}$
involving $C_{ijk\ell}$ 
couplings.}
\label{two-loop2}
\end{figure}
\end{center}

\begin{center}
\begin{figure}[htp]
\begin{picture}(410,140)(0,0)
\SetWidth{1.2}
{
\ArrowLine(115,70)(145,70)
\ArrowLine(145,70)(185,70)
\ArrowLine(205,70)(185,70)
\ArrowLine(225,70)(205,70)
\ArrowLine(225,70)(265,70)
\ArrowLine(265,70)(295,70)
\DashArrowArc(185,70)(40,0,90){5}
\DashArrowArc(185,70)(40,90,180){5}
\DashArrowArc(225,70)(40,180,270){5}
\DashArrowArc(225,70)(40,270,360){5}
\DashArrowLine(165,125)(185,110){5}
\DashArrowLine(185,110)(205,125){5}
\DashArrowLine(205,15)(225,30){5}
\DashArrowLine(225,30)(245,15){5}
\Text(115,65)[rt]{$q_L$}
\Text(280,65)[lt]{$q_R$}
\Text(160,65)[t]{$l$}
\Text(195,65)[t]{$q_R$}
\Text(215,65)[t]{$q_L$}
\Text(245,65)[t]{$l$}
\Text(205,70)[]{$\times $}
\Text(205,80)[t]{$m_q $}
\Text(140,90)[b]{$\phi_R^1$ }
\Text(235,90)[b]{$\phi_R^2$ }
\Text(215,125)[b]{$\phi_\ell^j$ }
\Text(155,125)[b]{$\phi_\ell^i$ }
\Text(185,36)[b]{$\phi_L^1$ }
\Text(270,36)[b]{$\phi_L^2$ }
\Text(195,10)[b]{$\phi_\ell^i$ }
\Text(260,10)[b]{$\phi_\ell^j$ }}
\end{picture}\\
\caption{ A further two-loop contribution
to $\bar{\theta}$}
\label{two-loop3}
%\end{figure}
%\end{center}

%\begin{center}
%\begin{figure}[ht]
\begin{picture}(410,140)(0,0)
\SetWidth{1.2}
{
\ArrowLine(125,20)(155,20)
\ArrowLine(155,20)(205,20)
\ArrowLine(205,20)(255,20)
\ArrowLine(255,20)(285,20)
\DashArrowArc(205,20)(50,90,180){5}
\DashArrowArcn(205,20)(50,90,0){5}
\DashArrowLine(205,90)(205,70){5}
\DashArrowLine(205,20)(205,70){5}
\Text(125,16)[rt]{$q_L$}
\Text(280,16)[lt]{$q_R$}
\Text(180,16)[t]{$l_R$}
\Text(230,16)[t]{$q_L$}
\Text(150,40)[b]{$\phi_R^1$ }
\Text(260,40)[b]{$\phi_\ell^i$ }
\Text(190,40)[b]{$\phi_R^2$ }
\Text(205,100)[b]{$\phi_\ell^j$ }}
\end{picture}\\
\caption{ An even further two-loop contribution to $\bar{\theta}$ }
\label{two-loop4}
%\end{figure}
%\end{center}

%\begin{center}
%\begin{figure}[ht]
\begin{picture}(410,140)(0,0)
\SetWidth{1.2}
{
\ArrowLine(125,20)(155,20)
\ArrowLine(155,20)(180,20)
\ArrowLine(180,20)(205,20)
\ArrowLine(205,20)(230,20)
\ArrowLine(230,20)(255,20)
\ArrowLine(255,20)(285,20)
\DashArrowArc(205,20)(50,90,180){5}
\DashArrowArcn(205,20)(50,90,0){5}
\DashArrowLine(205,20)(205,70){5}
\Text(125,16)[rt]{$q_L$}
\Text(280,16)[lt]{$q_R$}
\Text(180,16)[t]{$m_\ell$}
\Text(230,20)[]{$\times$}
\Text(180,20)[]{$\times$}
\Text(230,16)[t]{$m_\ell$}
\Text(150,40)[b]{$\phi_R^1$ }
\Text(260,40)[b]{$\phi_L^2$ }
\Text(190,40)[b]{$\phi_\ell^j$ }}
\end{picture}\\
\caption{ A final example of a two-loop contribution
to $\bar{\theta}$}
\label{two-loop5}
\end{figure}
\end{center}

\section{Discussion}

\bigskip
\bigskip

Because of the observation of large CP asymmetries
in decay of $(B^0, \bar{B}^0)$ at B Factories\cite{BF1,BF2}
the option of soft CP violation\cite{soft1,soft2,soft3}
as solution of strong CP is disfavored though not excluded.
Two popular alternative solutions, a massless up quark
and an (invisible) axion have theoretical and empirical
difficulties\cite{KKH++,lattice}.

\bigskip

It is therefore of interest to
find simple extensions which can solve strong CP. In the
present article we have shown how
unification
in a group $SU(3)^3 \times S_3$ leads to
a natural {\bf P} parity
operation and to sufficient
suppression of $\bar{\theta}$ provided 
CP is a high-energy symmetry and the
hierarchy of symmetry-breaking scales is in place.

\section*{Acknowledgments}
This work was supported in part by the US Department of Energy
under Grant No. DE-FG02-97ER-41036.

\newpage

\section*{Appendices}

\bigskip
\bigskip

\noindent{\large \bf Appendix A. ~~ The Stationary conditions.}\\

\bigskip
\bigskip

\noindent
The stationary conditions for each VEVs are 
\bea
& &v [m_{11}^2 + 2 (\lambda_{1111}+\eta_{1111}) v^2 
 + ( \lambda_{1122} + \lambda_{2211} +\eta_{1221} + \eta_{2112} ) w^2 
      \nn \\
& &~~~~~~~~~~~~~~~~~~~~~~~~~~~~~~~~
+ 2 \lambda_{1111} u_2^2) + (\lambda_{1122} + \lambda_{2211} ) u_3^2] 
+ 6 \gamma_{112} u_2 u_3 = 0
\label{condition v}\\
& & m_{22}^2 + 2 (\lambda_{2222}+\eta_{2222}) w^2 
 + ( \lambda_{1122} + \lambda_{2211} +\eta_{1221} + \eta_{2112} ) v^2 \nn \\
& &~~~~~~~~~~~~~~~~~~~~~~~~~~~~~~~~~~~~~~~~~~~~~~~~
+ (\lambda_{1122} + \lambda_{2211}) u_2^2 + 2 \lambda_{2222} u_3^2 = 0 
\label{condition w}\\
& &u_2 [m_{11}^2 + 2 \lambda_{1111} v^2 
+ (\lambda_{1122} + \lambda_{2211} )w^2 \nn \\
& &~~~~~~~~~~~~~~~~~~~~~~~
+ 2 (\lambda_{1111} + \eta_{1111} ) u_2^2 + (\lambda_{1122} + \lambda_{2211} ) u_3^2 ]
+ 6 \gamma_{112} u_3 v = 0 \\
& &u_3 [m_{22}^2 + (\lambda_{1122} + \lambda_{2211} )v^2 + 2 \eta_{2222} w^2 \nn \\
& &~~~~~~~~~~~~~~~~~~~~~~~
+  (\lambda_{1122} + \lambda_{2211} ) u_2^2 + 2 (\lambda_{2222} +\eta_{2222} ) u_3^2 ]
+ 6 \gamma_{112} u_2 v = 0 
\eea
At $m_{11},v \gg w \gg u_2, u_3$, by neglecting 
the terms for $w$ and $u_2,u_3$ in Eq.(\ref{condition v}), the size
of $v$ is   
\bea
v^2 = - \frac{m_{11}^2}{2 ( \lambda_{1111} + \eta_{1111} )} 
\eea 
Then, Eq.(\ref{condition w}) is 
\bea
m_{22}^2 + 2 (\lambda_{2222}+\eta_{2222}) w^2 
 + ( \lambda_{1122} + \lambda_{2211} +\eta_{1221} + \eta_{2112} ) v^2 
         = 0
\eea 
Hence, to realize the hierarchy of the scales $v >> w $, 
there are two possibilities. One is the terms of scale $w^2$ remain after 
cancelling between the terms of $v^2$ and $m_{22}^2$. The other one is 
the $m_{22}$ is the quantity of order $w$ and 
the size of the coefficients satisfy 
a condition as following,
\bea
& &\lambda_{1122} + \lambda_{2211} +\eta_{1221} + \eta_{2112} \leq 
   \left(\frac{w}{v} \right)^2, \\
& &\lambda_{iiii} + \eta_{iiii} \sim O(1),
\eea 
where $i=1,2$. 
These are the coefficients in the potential of $\phi_\ell $, the magnitude 
will correspond to the coefficients $A_{ijkl}^{L,R}$, $B_{ijkl}^{L,R}$ and
$C_{ijkl}$ of Eq.(\ref{CPV}) by $S_3$ symmetry.  

If we can assume that these coefficients of the mixing terms 
between $\phi^1$ and $\phi^2$ are suppressed by the factor, $\bar{\theta}$ 
from loop diagrams is also suppressed by the smallness of the coefficients,
because the imaginary parts come only from such mixing interactions by
the feature of the coefficients of Eq.(\ref{CPV}). \\   

\newpage

\bigskip
\bigskip 

\noindent{\large \bf Appendix B. ~~ Another option to reduce $\bar{\theta}$. }\\

\bigskip
\bigskip

\noindent
In the text, we assumed VEVs of $\phi_{\ell}^i$ as in Eq.(\ref{VEV}) 
and then found the
size of $\bar{\theta} $ will be smaller than $10^{-11}$. 
We mention another option to reduce $\bar{\theta} $. 
We may choose alternatively the set of the VEVs as follows:
\begin{equation}
\left\langle \phi_{\ell}^1 \right\rangle = \left( \begin{array}{ccc}
u_1 & 0 & 0 \\ 0&u_2 &  0 \\ 
0 & 0 & v
\end{array}
\right)  ~~ \hbox{and} ~~ 
\left\langle \phi_{\ell}^2 \right\rangle = \left( \begin{array}{ccc}
0  & 0 & 0 \\ 0& 0 & 0 \\ 
0 & w & 0  
\end{array}
\right) 
\label{VEV2}
\end{equation}  
In this case, we have only one constraint to the Yukawa couplings 
because the quarks and leptons get mass only from the VEV of $\phi_1$. 
So the Yukawa couplings are 
\bea
f_{1AB} &=& \frac{m_u^A}{u_1} \\
g_{1AB} &=& \frac{m_l^A}{u_2} 
\eea 
If we set all other Yukawa couplings to zero, 
the contributions
of all loop diagrams discussed in the text to $\bar{\theta} $
disappear because, by the relations among the coefficients in Eq.(\ref{CPV}), 
at least one coupling to the quark line in these graphs must 
be one of $\phi_2$, $f_{2AB}, g_{2AB}$. Hence, if there are 
no such couplings, an imaginary part of quark mass appears
only in one-loop diagrams with additional
propagators, leading to a suppression
of at least $\left( \frac{w}{v} \right)^4 \sim 10^{-14}$.

\end{document}